# Characteristics of Extinction Behavior of Ammonia Partial Cracking Simulated Fuels in Counterflow Premixed Flames under Lean Condition

**Jinseong Kim[a], Hyunchang Lee[a], Keeman Lee[a,*], Assaad R. Masri[b,*]**

[a]School of Mechanical and Aerospace Engineering/Center for Aerospace Engineering Research, Sunchon National University, Suncheon 57922, Republic of Korea
[b]School of Aerospace, Mechanical and Mechatronic Engineering, The University of Sydney, NSW 2006, Australia

## Abstract

In this study, the lean extinction behaviors of $CH_4$ and a 16% cracked ammonia surrogate fuel, CR16 (72% $NH_3$/21% $H_2$/7% $N_2$), in counterflow premixed flames were compared. Experiments were conducted along with one-dimensional simulations using OPPDIF; the GRI-Mech 3.0 and UCSD mechanisms were applied to $CH_4$ and CR16, respectively. CR16 exhibited higher extinction strain rates and resilience to strain-induced blow-out (RSIB) than those of $CH_4$ under lean conditions. At $\Phi = 0.66$, although both flames had the same extinction strain rate, CR16 showed a higher RSIB because of its longer flame time. Near its extinction, CR16 sustained $H_2$ consumption, radical reactions, and heat release through the interflame region toward the stagnation plane. Hydrogen oxidation coupled with nitrogen-related pathways, including $NH_2$ + NO and HNO reactions, supported this behavior. These results clarify the chemical basis for the superior extinction resistance of economically favorable low-cracking-ratio ammonia fuel.

*Corresponding authors.
E-mail addresses: kmlee@scnu.ac.kr (Keeman Lee), assaad.masri@sydney.edu.au (Assaad Masri).

## 1. Introduction

Ammonia has attracted considerable attention as a promising hydrogen carrier and carbon-free fuel because of its high hydrogen storage density and ease of storage and transportation [1–4]. However, ammonia has limitations in achieving stable ignition and flame stabilization in premixed combustion systems because of its high autoignition temperature, narrow range of flammability, low laminar-flame speed, and long ignition delay time. In addition, substantial fuel-$NO_x$ emissions may be generated owing to the nitrogen contained in the fuel. Therefore, both combustion stability and emission characteristics should be considered when ammonia is directly used as a fuel.

To overcome these limitations, extensive studies have been conducted on ammonia co-firing with highly reactive fuels such as hydrogen or methane [5–9]. In particular, the addition of hydrogen can improve flame stability by enhancing the flame speed and chemical reactivity. $NH_3/H_2/N_2$ mixed fuels produced through the partial cracking of ammonia are a promising approach because they can provide these benefits without requiring separate hydrocarbon-based auxiliary fuels, such as hydrogen or methane. Therefore, understanding the stretch-induced extinction behavior of partially cracked ammonia fuels is important for developing ammonia combustion systems with stable and low-emission performance.

Ammonia decomposition begins at approximately 200 °C, and the cracking efficiency varies depending on the application of a catalyst and the operating conditions. Cracking efficiencies of approximately 98–99% can be achieved at around 425 °C under catalytic conditions and at around 500 °C under non-catalytic conditions [10–12]. Following partial cracking, an $NH_3/H_2/N_2$ blended fuel is formed, consisting of unreacted residual $NH_3$ along with $H_2$ and $N_2$ produced through thermal decomposition. Recently, extensive research has been conducted on the fundamental combustion characteristics of these blended fuels [13–15]. A partially cracked ammonia surrogate fuel with a volumetric composition of 40% $NH_3$, 45% $H_2$, and 15% $N_2$, corresponding to a cracking ratio of 43%, was reported to exhibit combustion characteristics similar to those of conventional $CH_4$ fuel in terms of the flame speed and flame thickness in freely propagating laminar flames [16,17]. Jiang et al. [18] proposed a nitrogen-based reaction mechanism for the simulation of ammonia and ammonia/hydrogen flames and compared the combustion characteristics of the same $NH_3/H_2/N_2$ blended fuel with those of $CH_4$ flames. Their results showed that the $NH_3/H_2/N_2$ blended fuel could sustain counterflow premixed flames at higher strain rates than that sustained by $CH_4$ flames. Subsequently, Wiseman et al. [19] compared the blow-out behaviors of the same $NH_3/H_2/N_2$ blended fuel and $CH_4$ in turbulent premixed flames. They reported that the partially cracked ammonia surrogate fuel could maintain a flame even under higher flow-velocity conditions, demonstrating enhanced resistance to blow-out. Richter et al. [20] experimentally and numerically investigated the global and local extinction strain rates of 72% $NH_3$/21% $H_2$/7% $N_2$ (16% cracking ratio) and 64% $NH_3$/27% $H_2$/9% $N_2$ (22% cracking ratio) blended fuels in counterflow premixed flames. They also compared their resistance to extinction with that of $CH_4$ flames by using the dimensionless parameter called resilience to strain-induced blow-out (RSIB), defined as the product of the extinction strain rate and flame time. Their results showed that both blended fuels with cracking ratios of 16% and 22% exhibited higher RSIB values than those of $CH_4$ under lean conditions. In addition, the RSIB remained high in the lean regime over cracking ratios of 16%–28%. More recently, Xu et al. [21] investigated the non-monotonic dependence of extinction resilience on the ammonia cracking ratio using one-dimensional counterflow-flame simulations with modified transport models. They demonstrated that preferential diffusion plays an important role in the enhanced resilience at intermediate cracking ratios and significantly modifies the local $H_2$-consumption and flame structures.

However, direct comparisons between low-cracking-ratio $NH_3/H_2/N_2$ blended fuels and $CH_4$ flames under identical stretch conditions remain limited; The remaining question is not the cracking-ratio dependence itself, but how the enhanced extinction resistance of a representative low-cracking-ratio fuel is manifested in its detailed chemical structure relative to a conventional hydrocarbon flame under controlled strain conditions. Previous studies did not provide a direct CR16–$CH_4$ comparison of the spatially resolved species and radical production–consumption structures together with reaction-specific heat-release contributions near the stagnation plane. In particular, the coupling between $H_2$ oxidation and $NH_x$/HNO/NO-related chemistry sustaining the interflame reaction zone has not been addressed in such a direct comparison. In addition, experimental studies comparing the extinction limits of low-cracking-ratio $NH_3/H_2/N_2$ blended fuels and $CH_4$ flames, together with detailed analyses of the heat-release and chemical-reaction structures in the interflame region under the same global strain-rate conditions, are scarce. Quantitative comparisons of the relationship between the enhanced resistance to extinction and hydrogen-consumption reactions sustained near the stagnation plane, heat-release distributions, and nitrogen-related reaction pathways are essential for improving the understanding of the extinction resistance of ammonia-cracking fuels.

Therefore, this study aimed to provide an in-depth investigation of the extinction resistance of an $NH_3/H_2/N_2$ surrogate fuel with a 16% cracking ratio (72% $NH_3$/21% $H_2$/7% $N_2$), which exhibits a high RSIB value under lean conditions while requiring a relatively low cracking-conversion energy, from an economic perspective. First, the extinction behavior of lean counterflow premixed flames fueled by the 16% cracked $NH_3/H_2/N_2$ surrogate fuel was experimentally and numerically compared with that of $CH_4$, the main constituent of conventional natural gas (NG). To this end, the extinction strain rates predicted by the ammonia-related chemical kinetic mechanisms were first validated against experimental measurements, and an appropriate mechanism for $NH_3/H_2/N_2$ flame simulations was selected. Subsequently, the extinction strain rate and RSIB of the two fuels were compared over the equivalence-ratio range of $\Phi$ = 0.5–0.7, which is relevant to the operating conditions of conventional combustion devices, such as gas turbines. In particular, detailed analyses of the temperature, heat-release rate (HRR), formation and consumption reactions of major chemical species, and reaction-specific heat-release contributions were conducted at $\Phi = 0.66$, at which the extinction strain rates of the two flames were identical. Through these analyses, the heat release sustained throughout the interflame region and near the stagnation plane, together with the contributions of hydrogen oxidation and major chemical reactions in the 16% cracked fuel flame (hereinafter referred to as CR16), were examined to clarify how hydrogen-related reactions are coupled with ammonia-specific nitrogen chemistry and contribute to the detailed reaction and heat-release structures associated with the enhanced extinction resistance of the low-cracking-ratio ammonia fuel.

## 2. Experimental setup & numerical simulation method

### *2.1. Experimental setup*

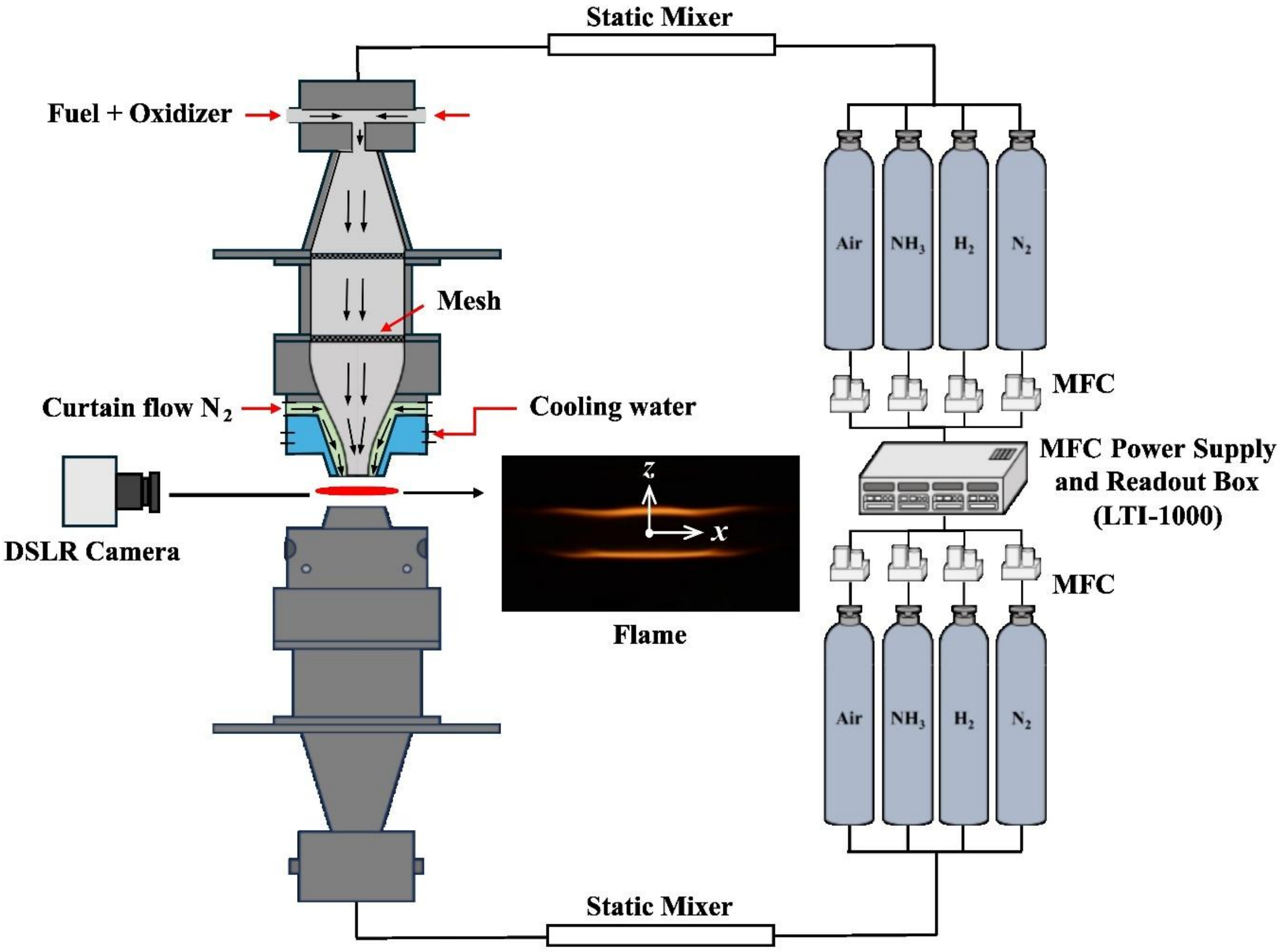

**Fig. 1.** Schematic of the experimental setup for the counterflow burner.

A schematic of the experimental apparatus for the counterflow premixed flame analyzed in this study is shown in Fig. 1. The counterflow burner consists of vertically opposed, coaxial double nozzles; the premixed fuel/oxidizer mixtures are issued from the upper and lower main nozzles. The flow rates of the fuel and oxidizer were precisely controlled using mass flow controllers (MFCs; Lintec, accuracy: ±1.0% of full scale) and an MFC Power Supply and Readout Box (LTI-1000). The full-scale flow rates of the MFCs used for $NH_3$, $H_2$, $N_2$, and air were 3 slpm, 2 slpm, 600 sccm, and 20 slpm, respectively. Static mixers were installed in each flow line to promote uniform mixing of the fuel and oxidizer, and meshes placed inside the nozzles were used to establish uniform velocity profiles at the nozzle exits.

Each nozzle consisted of a central main nozzle and an outer nitrogen curtain nozzle. Premixed reactants were supplied through the main nozzle, whereas nitrogen was introduced through the outer nozzle as a curtain flow. The exit velocity of the nitrogen curtain flow was set equal to the bulk velocity of the main-nozzle mixture to minimize the entrainment of ambient air, outer wake flow, and flow disturbances caused by the shear layer. The diameter of the main nozzle and separation distance between the upper and lower nozzles were 10 and 12 mm, respectively. The purities of $CH_4$, $NH_3$, $H_2$, and $N_2$ used to prepare the partially cracked ammonia surrogate fuel were all 99.999%. Direct flame images were acquired using a DSLR camera (Nikon D3200). The $NH_3/H_2/N_2$ blended fuel used in this study was formulated to simulate partially cracked ammonia. The ammonia-cracking ratio and corresponding chemical reaction as a function of the cracking ratio are given in Eqs. (1) and (2), respectively [17].

$$\alpha = \frac{Y_{H_2} + Y_{N_2}}{Y_{NH_3} + Y_{H_2} + Y_{N_2}} \qquad (1)$$

$$(1-\alpha)NH_3 + \alpha\left(\frac{3}{2}H_2 + \frac{1}{2}N_2\right) + \frac{3}{4}(O_2 + 3.76N_2) \rightarrow \frac{3}{2}H_2O + 3.32N_2 \qquad (2)$$

In Eq. (1), $Y_i$ denotes the mass fraction of the species $i$ in the fuel mixture. A previous study [20] reported that partially cracked ammonia surrogate fuels with cracking ratios ranging from 16% to 28% exhibited higher RSIB values than that of methane under lean combustion conditions, whereas lower cracking ratios required less ammonia-cracking conversion energy. Therefore, this study compared the extinction behavior of lean counterflow premixed flames fueled by a 16% cracked $NH_3/H_2/N_2$ blended fuel (72% $NH_3$/21% $H_2$/7% $N_2$), which maintains a high RSIB value in the lean regime while imposing the lowest ammonia-cracking conversion burden, with that of $CH_4$, the main constituent of NG. For each equivalence-ratio condition, the nozzle-exit velocity was increased in increments of 5 cm/s until flame extinction occurred, thereby establishing the extinction boundary. As the nozzle-exit velocity increased, the global strain rate experienced by the flames also increased. The global strain rate used in this study is defined by Eq. (3) [22].

$$a_g = \frac{2V_2}{L}\left(1 + \frac{V_1\sqrt{\rho_1}}{V_2\sqrt{\rho_2}}\right), \qquad (3)$$

where L denotes the separation distance between the upper and lower nozzles, which was 12 mm in the present study; V and $\rho$ represent the nozzle-exit velocity and density of the premixed reactants, respectively. The subscripts 1 and 2 denote the upper and lower nozzles, respectively. In this study, a symmetric counterflow premixed flame configuration was employed, in which premixed reactants of identical composition were supplied to the upper and lower nozzles at the same velocity. Therefore, by applying $V = V_1 = V_2$ and $\rho = \rho_1 = \rho_2$ to the momentum-conservation relation $V_1\sqrt{\rho_1} = V_2\sqrt{\rho_2}$, the expression for the global strain rate can be simplified to $a_g = 4V/L$.

## *2.2. Numerical simulation method*

Numerical simulations were performed using the OPPDIF code [23] implemented in ANSYS Chemkin-Pro. The OPPDIF code solves the flame established between opposed nozzles as a one-dimensional problem and is suitable for analyzing the temperature, species concentration, HRR, and velocity distribution of counterflow premixed flames. The initial pressure and temperature at both the upper and lower nozzles were set to 1 atm and 298 K, respectively. To simulate a symmetric counterflow premixed flame, premixed reactants with identical compositions were supplied to both nozzles at the same inlet velocity. The nozzle separation distance was set to 12 mm, consistent with the experimental configuration. To assess the sensitivity of the predicted extinction behavior to molecular transport and to examine the influence of species-dependent diffusion, calculations were performed using mixture-averaged, multicomponent, and unity-Lewis-number transport models. The mixture-averaged model was used as the baseline transport model, while the multicomponent model was employed to evaluate the sensitivity to a more detailed treatment of species diffusion. In the unity-Lewis-number calculations, the Lewis number of all species was set to unity ($Le_i = 1$), thereby suppressing species-dependent differential diffusion. Thermal diffusion (Soret diffusion) was included in the mixture-averaged and multicomponent calculations. To resolve the temperature, HRR, and species-concentration profiles of the counterflow premixed flames in detail, the computational grid was adaptively refined to contain approximately 600–900 points, with 900 points set as the maximum allowable grid size. To assess the grid sensitivity of the predicted global extinction strain rates, additional calculations were performed using three adaptive-grid ranges: approximately 300–600, 600–900, and 900–1200 points over the primary equivalence-ratio range of $\Phi = 0.5 - 0.7$, where the upper value in each range denotes the maximum allowable number of grid points. For the $CH_4$ flames, identical global extinction strain rates were obtained for all three grid resolutions over the investigated equivalence-ratio range. For the CR16 flames, the extinction strain rates were also insensitive to further grid refinement, except for minor differences at $\Phi = 0.65$ and $0.70$. At $\Phi = 0.65$, the extinction strain rates were 933 $s^{-1}$ for the 300-600- and 600-900-point grids and 950 $s^{-1}$ for the 900-1200-point grid. Similarly, at $\Phi = 0.70$, values of 1000 and 1017 $s^{-1}$ were obtained for the 300-600/600-900-point and 900-1200-point grids, respectively. These differences are approximately 1.7–1.8%. The experimental and numerical conditions for the $CH_4$ and $NH_3/H_2/N_2$ blended-fuel flames are summarized in Table 1. The $NH_3/H_2/N_2$ blended fuel corresponds to an ammonia-cracking ratio of 16%; hereinafter, this fuel is referred to as CR16.

**Table 1.** Experimental and numerical conditions for the $CH_4$ and CR16 flames.

| Parameter | Unit | $CH_4$ | CR16 |
|---|---|---|---|
| Global strain rate, $a_g$ | $s^{-1}$ | 10 – 1217 | 10 – 1000 |
| Composition (Vol. %) | - | 100% $CH_4$ | 72% $NH_3$/ 21% $H_2$/ 7% $N_2$ |
| Equivalence ratio, Φ | - | 0.5 – 0.7 | |
| Distance of nozzle, L | mm | 12 | |

## *2.3. Mechanism verification*

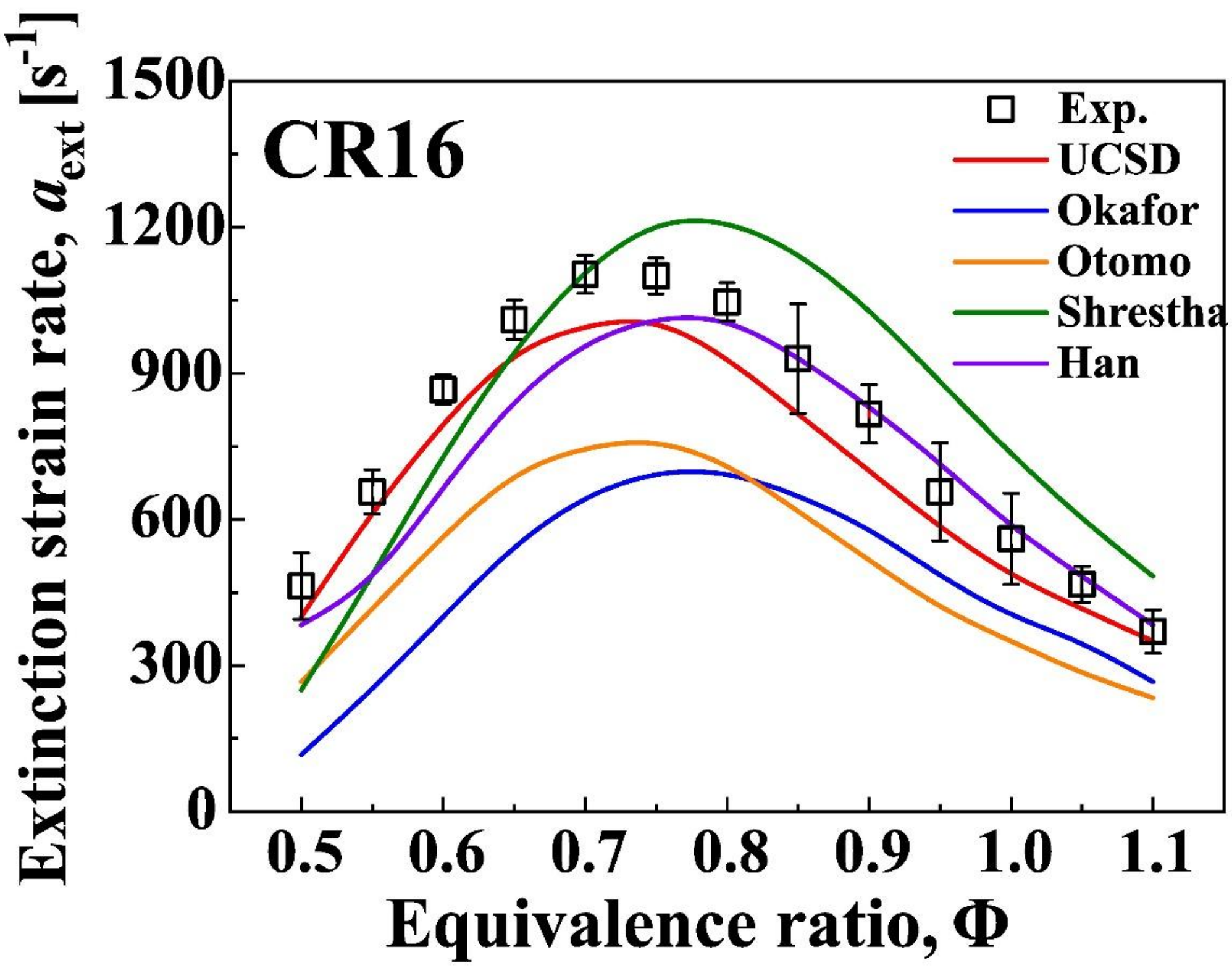

**Fig. 2.** Validation of chemical kinetic mechanisms against experimentally measured global extinction strain rates of CR16 counterflow flames using the mixture-averaged transport model.

GRI-Mech 3.0 [24] was employed for the numerical simulations of $CH_4$ flames. GRI-Mech 3.0 consists of 53 chemical species and 325 reactions and is widely used for the combustion simulations of $CH_4$ and natural-gas-based fuels. In contrast, because CR16 is an $NH_3$/$H_2$/$N_2$ blended fuel involving nitrogen-related reaction pathways, a suitable kinetic mechanism capable of appropriately predicting the experimental extinction boundaries was selected prior to the numerical analysis. To this end, the global extinction strain rates calculated using the UCSD [25], Okafor [26], Otomo [27], Shrestha [28], and Han [29] mechanisms were compared with the experimental results. For this mechanism-validation procedure, the mixture-averaged transport model was employed as the baseline transport model for all five kinetic mechanisms.

Fig. 2 shows the comparison of the experimentally measured and numerically predicted global extinction strain rates of CR16 counterflow premixed flames with those predicted using the five kinetic mechanisms under the mixture-averaged transport model. Here, the global extinction strain rate is defined as the global strain rate of the flame condition immediately before extinction. The experiments were repeated at least five times for each equivalence-ratio condition. The extinction point was determined by increasing the nozzle-exit velocity in increments of 5 cm/s. The uncertainty of the experimentally determined global extinction strain rate was evaluated following standard uncertainty-analysis procedures [30,31], with consideration of the measurement repeatability, the ±1.0% full-scale accuracy of the MFCs, and the finite velocity increment used to identify the extinction boundary. A similar uncertainty-propagation approach has also been employed in previous counterflow extinction measurements [9]. The repeatability contribution was evaluated as $\mathrm{u_{rep}} = \mathrm{s}/\sqrt{\mathrm{n}}$, where sis the sample standard deviation of the five repeated measurements. Assuming rectangular distributions for the MFC accuracy and the extinction-search interval, the corresponding standard uncertainties were evaluated by dividing their half-widths by $\sqrt{3}$.

For the present nozzle separation of 12 mm, the velocity increment of 5 cm/s corresponds to a global strain-rate increment of approximately 16.7 $s^{-1}$. Therefore, the standard uncertainty associated with the finite extinction-search increment was $u_{\mathrm{step}} = (16.7/2)/\sqrt{3} = 4.81\ \mathrm{s}^{-1}$. Propagation of the ±1.0% full-scale uncertainties of the individual MFCs resulted in an MFC contribution of approximately $u_{\mathrm{MFC}} = 5.87\ \mathrm{s}^{-1}$ for the CR16 experiments. The combined standard uncertainty was calculated as

$$u_c = \sqrt{u_{\mathrm{rep}}^2 + u_{\mathrm{MFC}}^2 + u_{\mathrm{step}}^2} \qquad (4)$$

and the expanded uncertainty was evaluated using a coverage factor of $k = 2$, i.e., $U = 2u_c$. The resulting expanded uncertainties of the CR16 extinction strain rates ranged from approximately 15.2 to 56.2 $s^{-1}$, corresponding to relative uncertainties of approximately 1.75–8.35% over $\Phi = 0.5$–1.1. The error bars in Fig. 2 represent these expanded uncertainties.

In the numerical simulations, the inlet velocities at the upper and lower nozzles were likewise increased in increments of 5 cm/s to identify the extinction point, following the same procedure as in the experiments.

The Okafor and Otomo mechanisms predicted lower global extinction strain rates than the experimental values over most of the equivalence-ratio range. The Shrestha mechanism overpredicted the extinction strain rate under very lean conditions, whereas it predicted higher values than the experimental results at $\Phi \geq 0.75$. The Han mechanism relatively well reproduced the experimental trends over the equivalence-ratio range of 0.75–1.10; however, it underpredicted the extinction strain rate in the lean regime, which is the primary focus of the present study. The UCSD mechanism reproduced the experimental trends most consistently over the entire equivalence-ratio range and showed the closest agreement with the experimental results, particularly under lean conditions of $\Phi \leq 0.7$. The discrepancies between the experimentally measured and predicted extinction boundaries over the entire equivalence-ratio range were 9.55% and 10.06% for the UCSD and Han mechanisms, respectively. However, within the primary analysis range of the present study ($\Phi = 0.5$–0.7), the discrepancies for the UCSD and Han mechanisms were 8.55% and 19.55%, respectively. Therefore, the UCSD mechanism, which most accurately predicted the lean extinction boundary among the five mechanisms examined in this study, was adopted for the subsequent numerical analysis of CR16 flames.

# Results and discussion

## *3.1. Extinction behavior of the $CH_4$ and CR16 flames*

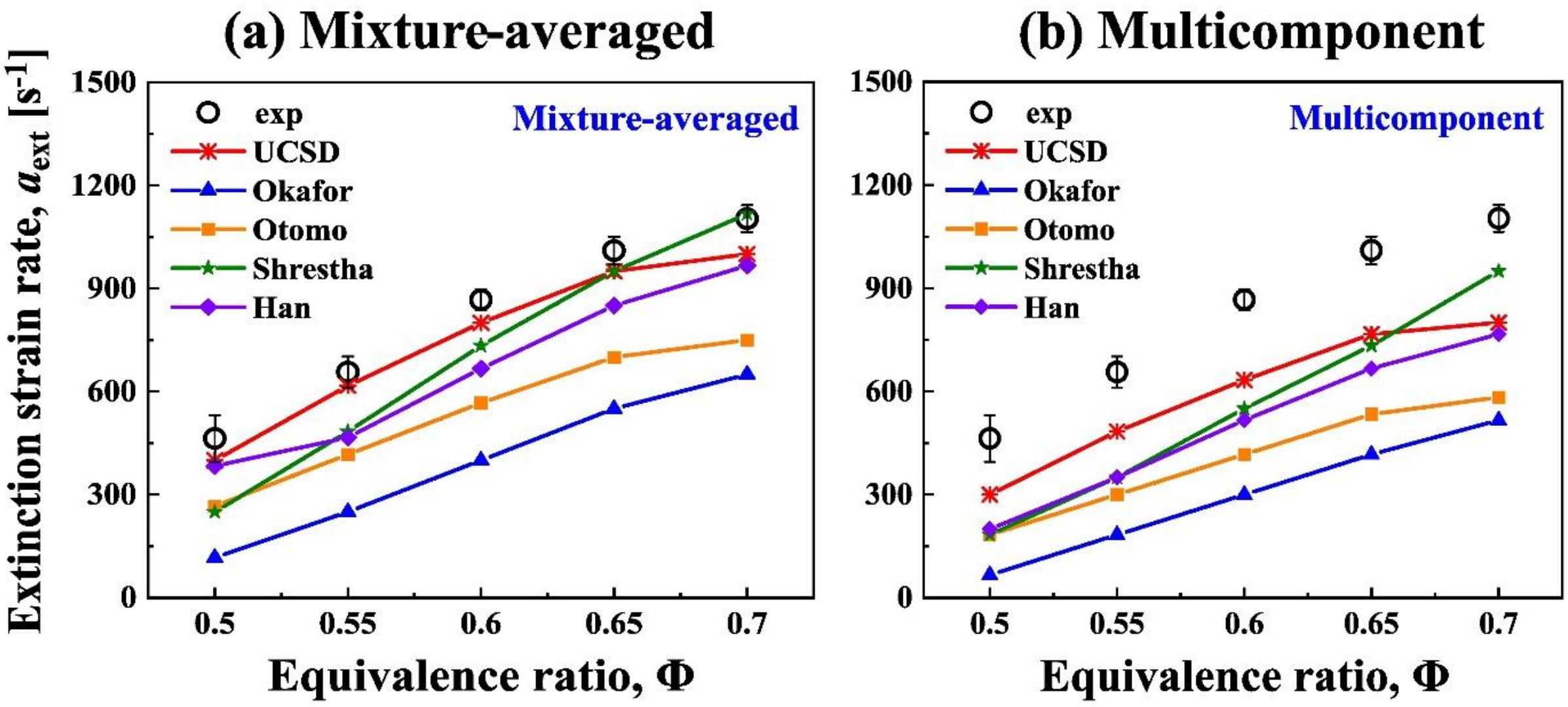

**Fig. 3.** Comparison of experimentally measured and numerically predicted global extinction strain rates of CR16 counterflow premixed flames using five chemical kinetic mechanisms with (a) mixture-averaged and (b) multicomponent transport models over the lean equivalence-ratio range of $\mathbf{\Phi = 0.5–0.7}$.

To further examine whether the predicted extinction behavior and the mechanism-selection results were sensitive to the treatment of molecular transport, additional calculations were performed using the multicomponent transport model over the primary analysis range of $\Phi =$ 0.5–0.7. Fig. 3 compares the extinction strain rates predicted using the five kinetic mechanisms with the mixture-averaged and multicomponent transport models. Compared with the mixture-averaged results, the multicomponent transport model predicted systematically lower extinction strain rates for all five mechanisms throughout the investigated equivalence-ratio range. Consequently, the mixture-averaged calculations showed closer agreement with the experimentally measured extinction boundaries under the present kinetic mechanisms.

Despite the quantitative differences in the predicted extinction strain rates, both transport models preserved the overall increase in extinction strain rate with increasing equivalence ratio and showed broadly similar trends among the kinetic mechanisms. Importantly, the UCSD mechanism exhibited the lowest discrepancy from the experimental extinction boundary under both transport models. Under the multicomponent model, the discrepancy for the UCSD mechanism was 28.03%, which remained the lowest among the five mechanisms examined. Therefore, although the absolute extinction strain rates were sensitive to the transport model, the mixture-averaged transport model combined with the UCSD mechanism was adopted as the baseline condition for the subsequent CR16 analyses.

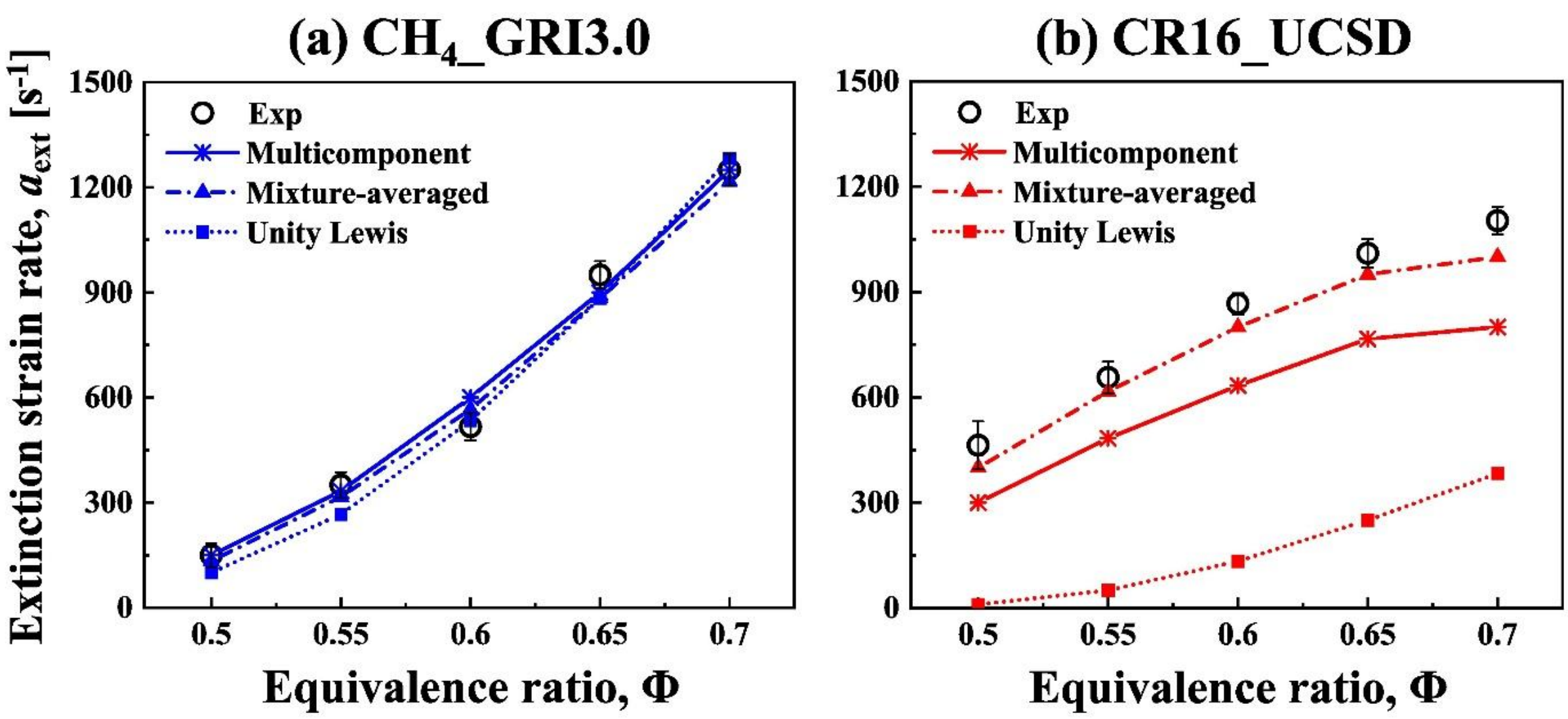

**Fig. 4.** Effects of transport model on the global extinction strain rates of (a) $CH_4$ flames using GRI-Mech 3.0 and (b) CR16 flames using the UCSD mechanism over $\mathbf{\Phi = 0.5}$–$\mathbf{0.7}$.

Based on the mechanism-validation and transport-sensitivity results, GRI-Mech 3.0 and UCSD were retained for the subsequent analyses of the $CH_4$ and CR16 flames, respectively. To further examine the role of species-dependent transport in the extinction behavior, additional calculations were performed using a unity-Lewis-number model, in which the Lewis numbers of all species were set to unity.

Fig. 4 compares the experimentally measured extinction strain rates with the predictions obtained using the multicomponent, mixture-averaged, and unity-Lewis-number transport models. For the $CH_4$ flames, the predicted extinction strain rates were only weakly affected by the transport model, with the three predictions remaining close over the investigated equivalence-ratio range. In contrast, the CR16 flames exhibited a pronounced sensitivity to the transport treatment. Although the mixture-averaged and multicomponent models preserved similar increasing trends with equivalence ratio, the multicomponent model predicted systematically lower extinction strain rates. More importantly, imposing unity Lewis numbers resulted in a substantial reduction in the extinction strain rate throughout the investigated range. This indicates that species-dependent transport plays an important role in their extinction resistance. In contrast, the comparatively small variation observed for the $CH_4$ flames suggests a weaker sensitivity to differential transport under the present lean conditions.

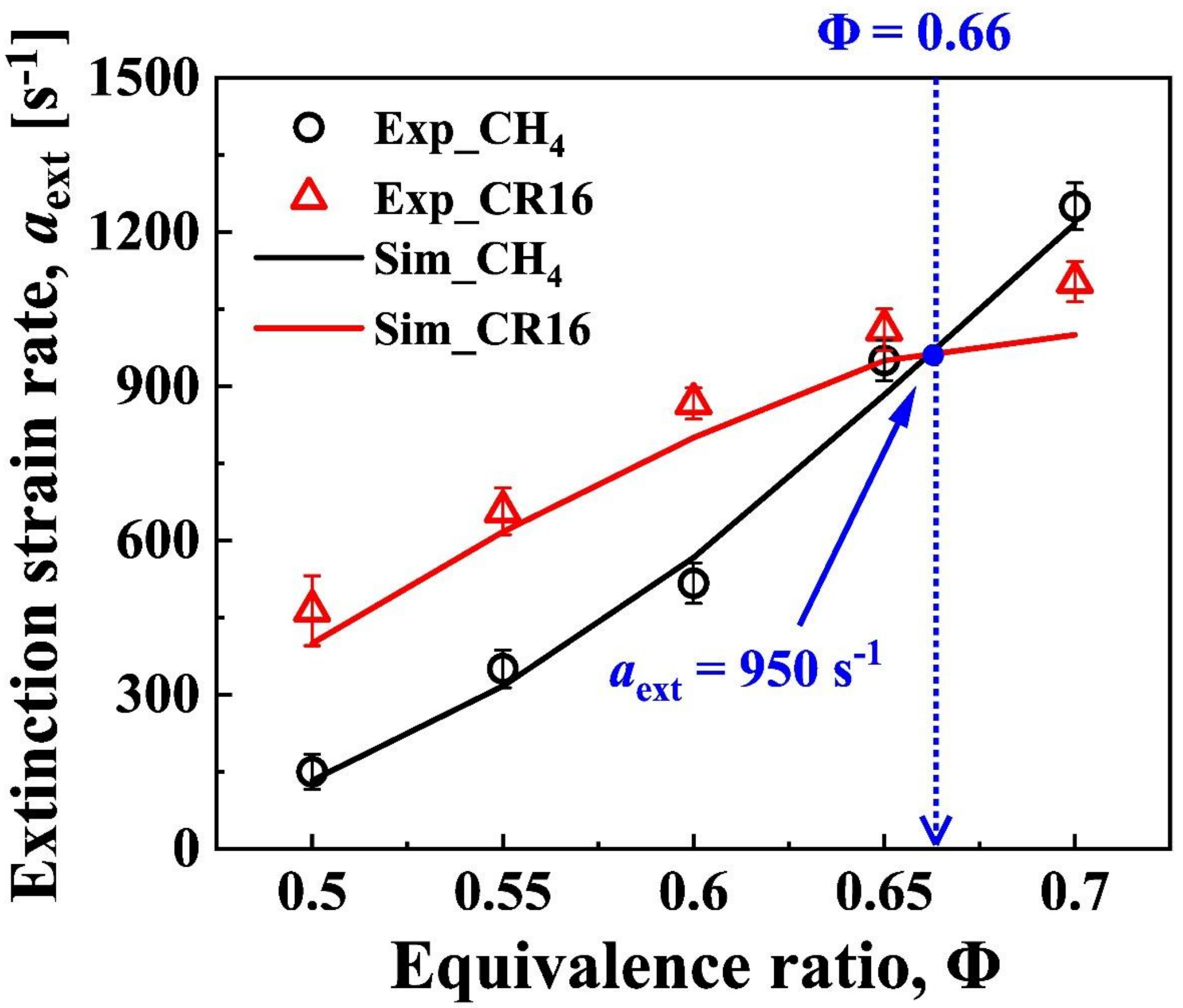

**Fig. 5.** Comparison of experimentally measured and simulated global extinction strain rates of $CH_4$ and CR16 counterflow premixed flames using the mixture-averaged transport model under lean conditions (**Φ** = 0.5–0.7).

Fig. 5 compares the experimentally measured and numerically predicted global extinction strain rates of $CH_4$ and CR16 flames using the baseline mixture-averaged transport model over the lean equivalence-ratio range of Φ = 0.5–0.7. The symbols and solid lines represent the experimental and numerical results, respectively. At Φ = 0.5, the CR16 flame exhibited a higher global extinction strain rate than that of the $CH_4$ flame, indicating that the CR16 flame could be stably sustained even under higher stretch conditions. As the equivalence ratio increased, the difference in the global extinction strain rate between the two fuels gradually decreased. Direct numerical calculations performed at Φ = 0.66 yielded the same global extinction strain rate of $a_g = 950\ s^{-1}$ for both the $CH_4$ and CR16 flames. Beyond Φ = 0.66, the global extinction strain rate of the CR16 flame increased gradually, whereas that of the $CH_4$ flame increased more rapidly with increasing equivalence ratio. Consequently, the trends of the extinction strain rates of the two fuels were reversed.

### *3.2. Characteristics of unstretched laminar flame*

Before conducting a detailed analysis of the extinction behavior of counterflow premixed flames, the fundamental combustion characteristics under freel y propagating flame conditions were compared. To this end, unstretched laminar freely propagating premixed flames of $CH_4$ and CR16 were simulated using the PREMIX code implemented in ANSYS Chemkin-Pro. The flame speed ($S_L$), flame thickness ($\delta_F$), and flame time ($t_F$) are related as shown in Eq. (5) [32]. These quantities were subsequently used, together with the extinction strain rate, to analyze the RSIB of the flames.

$$t_F = \frac{\delta_F}{S_L} = \frac{T_b - T_u}{|dT/dz|_{max} S_L} \tag{5}$$

In Eq. (5), $T_b$ and $T_u$ denote the temperatures of the burned and unburned gases, respectively, and $|dT/dz|_{max}$ represents the maximum axial temperature gradient. The flame thickness $\delta_F$ was determined based on the maximum temperature gradient, and the flame time $t_F$ was subsequently calculated by dividing $\delta_F$ by the laminar-flame speed $S_L$.

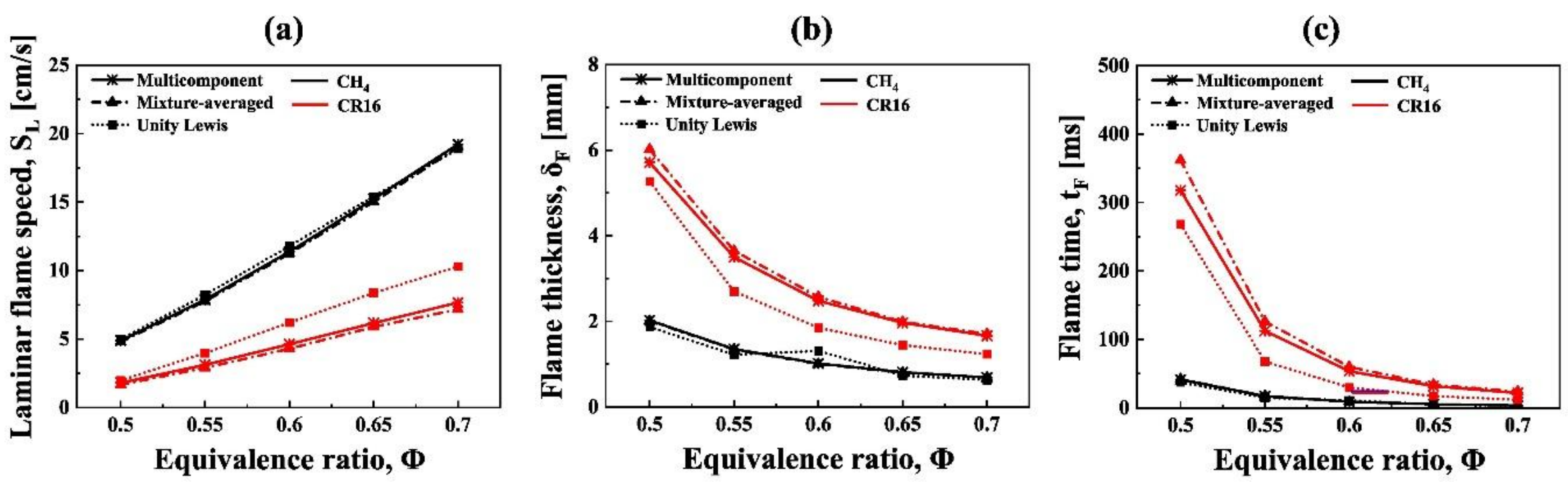

**Fig. 6.** Effects of transport model on the unstretched laminar-flame characteristics of $CH_4$ and CR16 premixed flames over $\mathbf{\Phi = 0.5–0.7}$: (a) laminar flame speed, $\mathbf{S_L}$, (b) flame thickness, $\boldsymbol{\delta_F}$, and (c) flame time, $\mathbf{t_F}$.

Fig. 6 shows the comparison of the laminar-flame speed, flame thickness, and time of free propagation of $CH_4$ and CR16 premixed flames as functions of the equivalence ratio using the mixture-averaged, multicomponent, and unity-Lewis-number transport models. For the $CH_4$ flames, the three transport models yielded similar values of $S_L$, $\delta_F$, and $t_F$ over the investigated equivalence-ratio range, indicating a relatively weak sensitivity of the unstretched flame characteristics to the treatment of molecular model. In contrast, the CR16 flames exhibited a more pronounced dependence on the transport model. Under both the mixture-averaged and multicomponent models, CR16 exhibited a lower laminar flame speed and a greater flame thickness than $CH_4$ over the entire equivalence-ratio range. Consequently, the flame time of CR16 was substantially longer than that of $CH_4$, particularly under very lean conditions. Although quantitative differences were observed between the mixture-averaged and multicomponent predictions, the relative characteristics of the two fuels were consistently maintained.

A more pronounced change was observed when the Lewis numbers of all species were set to unity. For CR16, the unity-Lewis-number model resulted in a higher laminar flame speed and a smaller flame thickness than those predicted using the mixture-averaged and multicomponent models, leading to a substantial reduction in flame time. In comparison, the corresponding changes in the $CH_4$ flame were relatively small. This contrast demonstrates that the unstretched flame characteristics of CR16 are considerably more sensitive to species-dependent transport effects. Because the unity-Lewis-number model suppresses both differential diffusion among species and the effects associated with Lewis numbers deviating from unity, the differences from the mixture-averaged and multicomponent results represent the combined influence of these transport effects. More generally, characteristic timescales are important in determining flame response, because a reduction in the time available for chemical reaction under increasing strain can limit reaction progress and ultimately promote flame extinction [33-35]. Richter et al. [20] introduced RSIB as the product of the extinction strain rate and laminar flame time to enable a dimensionless comparison of $NH_3/H_2/N_2$ and $CH_4$ flames, whereas Xu et al. [21] employed the equivalent definition in terms of the extinction Karlovitz number, $Ka_{\mathrm{ext}} = a_{\mathrm{ext}}\delta_{\mathrm{F}}/S_{\mathrm{L}}$. Accordingly, RSIB should be interpreted as a dimensionless extinction response normalized by the intrinsic flame timescale, rather than as a direct measure of the absolute strain rate that a flame can withstand. In this context, $t_{\mathrm{F}}$ represents an intrinsic flame timescale that can be compared with the flow timescale imposed by strain. The substantially longer flame time of CR16 under the mixture-averaged and multicomponent models contributes directly to its larger RSIB. However, a longer flame time alone does not necessarily indicate a greater resistance to stretch-induced extinction, because RSIB is determined by both the characteristic flame time and the local strain rate sustained immediately prior to extinction. Therefore, the extinction strain rate and RSIB are considered separately in the following analysis because the former characterizes the absolute strain level sustained before extinction, whereas the latter characterizes the extinction response relative to the intrinsic flame timescale.

Next, the RSIB values of the $CH_4$ and CR16 flames were compared using the flame time calculated by the PREMIX code and local extinction strain rate calculated by the OPPDIF code. RSIB is defined as the product of the flame time and local strain rate immediately prior to extinction. The RSIB and local strain rate are expressed by Eqs. (6) and (7), respectively.

$$\mathrm{RSIB} = t_{\mathrm{F}} \times a_{\mathrm{loc,\ ext}} \quad (6)$$

$$a_{\mathrm{loc}} = |du_z/dz|_{max} \quad (7)$$

In Eq. (6), $t_F$ denotes the flame time calculated using the PREMIX code, and $a_{loc,ext}$ represents the local strain rate calculated using the OPPDIF code under conditions immediately prior to extinction. In Eq. (7), $u_z$ and $z$ denote the axial velocity and axial coordinate along the $z$-direction shown in Fig. 1, respectively. The local strain rate is defined as the maximum absolute value of the axial velocity gradient within the computational domain.

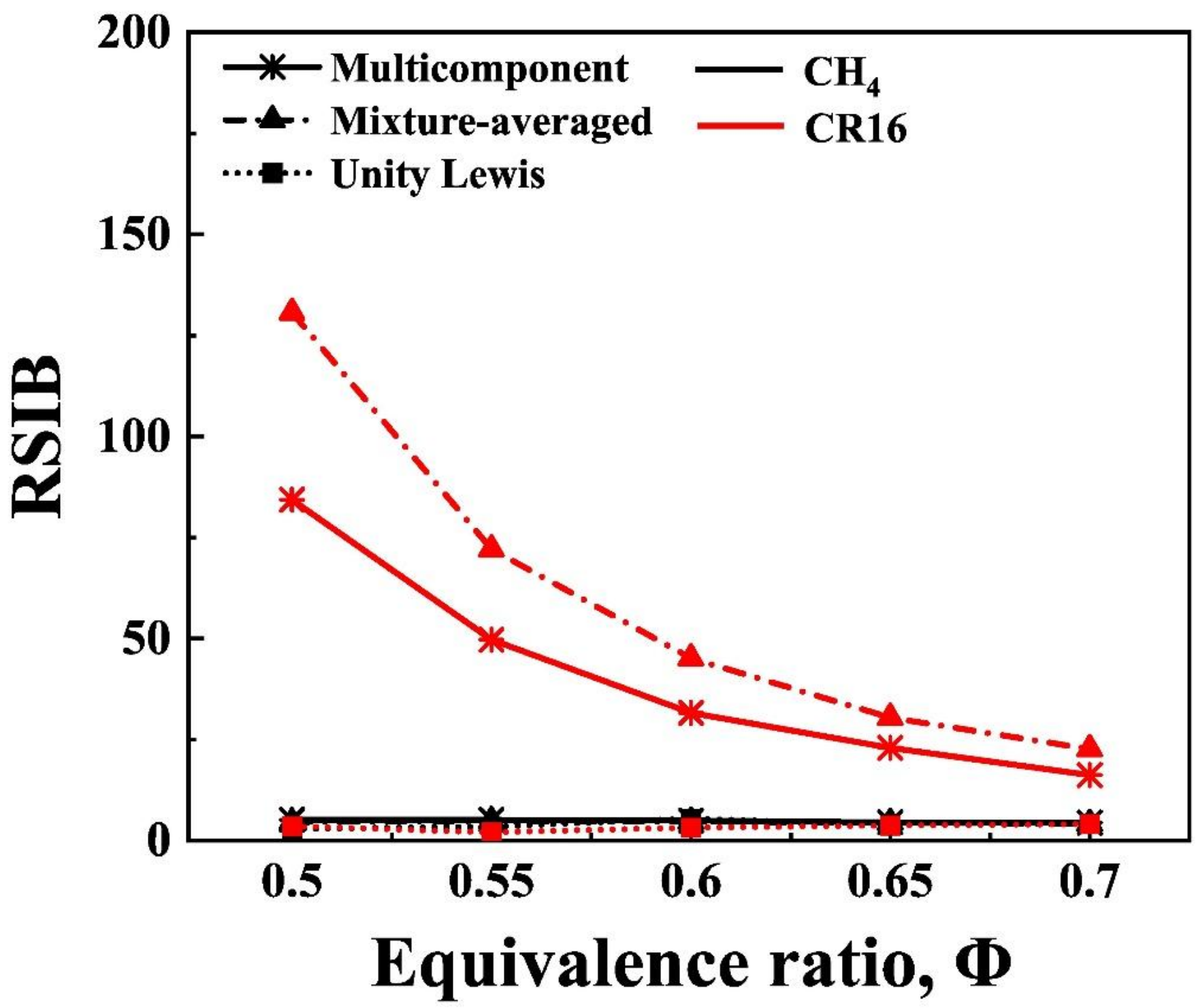

**Fig. 7.** Comparison of the resilience to strain-induced blow-out (RSIB) of $CH_4$ and CR16 counterflow premixed flames calculated using mixture-averaged, multicomponent, and unity-Lewis-number transport models over **Φ** = 0.5–0.7.

Fig. 7 compares the RSIB values of the $CH_4$ and CR16 flames as functions of the equivalence ratio for the mixture-averaged, multicomponent, and unity-Lewis-number transport models. For each transport model, the RSIB was calculated using the corresponding flame time obtained from PREMIX and the local extinction strain rate obtained from OPPDIF. The RSIB of the CR16 flame was markedly higher than that of the $CH_4$ flame under all lean conditions, and the difference between the two fuels became more pronounced as the equivalence ratio decreased. The larger RSIB of CR16 reflects the combined contributions of its longer intrinsic flame time and the local strain rate at extinction. Because CR16 has a lower laminar-flame speed and greater flame thickness than $CH_4$, its substantially longer flame time makes an important contribution to RSIB. Therefore, the larger RSIB should not, by itself, be interpreted as evidence of a higher absolute extinction strain rate. As shown in Fig. 5, the extinction strain rate of CR16 increased with increasing equivalence ratio, whereas its RSIB decreased. This is because the flame time decreased considerably with increasing equivalence ratio as the laminar-flame speed increased and flame thickness decreased; consequently, the reduction in flame time had a greater effect than the increase in the extinction strain rate. This opposite trend further demonstrates that the extinction strain rate and RSIB characterize distinct aspects of the flame response. The sensitivity of RSIB to the transport model was also markedly different between the two fuels. The $CH_4$ flame showed relatively weak dependence on the transport model, whereas CR16 exhibited a much stronger response, particularly under the unity-Lewis-number model. The substantial variation in the RSIB of CR16 when the Lewis numbers of all species were set to unity indicates that species-dependent transport plays an important role in its dimensionless extinction response. Because the unity-Lewis-number model suppresses both differential diffusion among species and the effects associated with Lewis numbers deviating from unity.

### *3.3. Temperature and HRR*

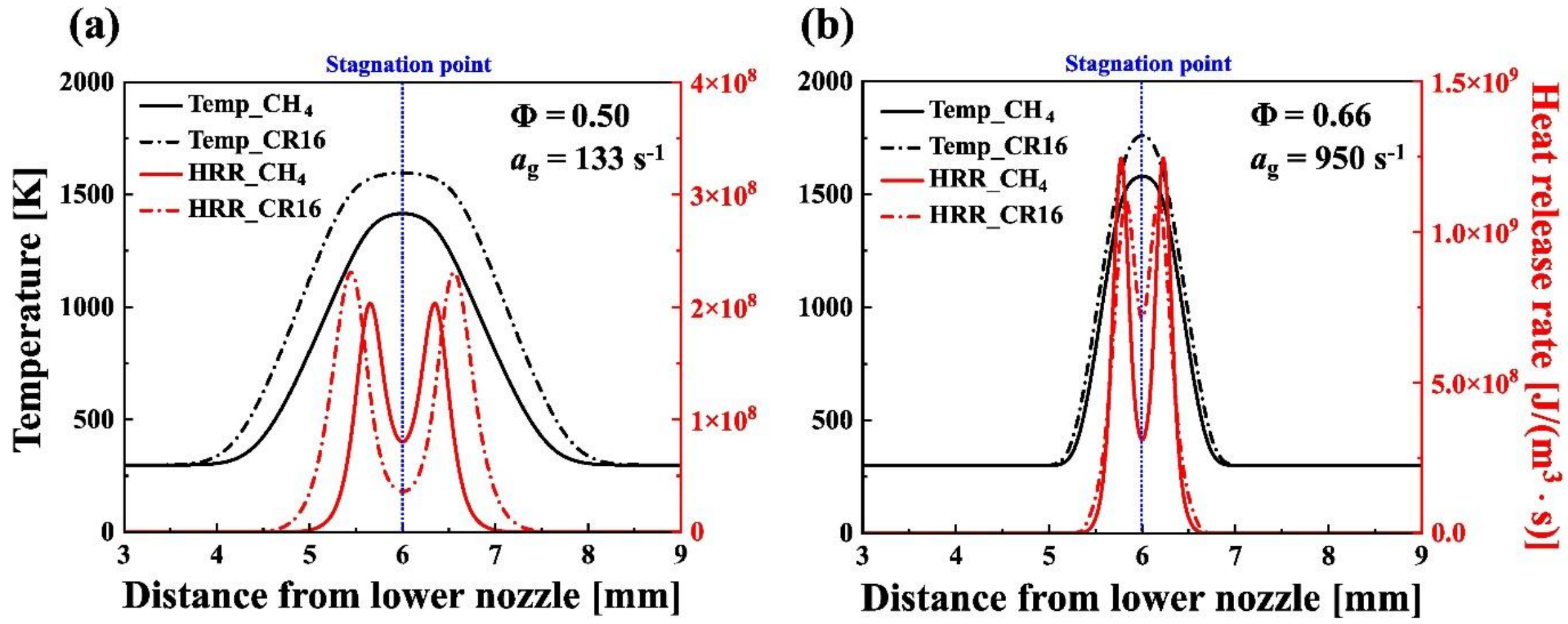

**Fig. 8.** Temperature and heat-release rate profiles of $CH_4$ and CR16 counterflow premixed flames at (a) $\boldsymbol{\Phi}$ = 0.5, $\boldsymbol{a}_{\mathbf{g}}$ = 133 $s^{-1}$ and (b) $\boldsymbol{\Phi}$ = 0.66, $\boldsymbol{a}_{\mathbf{g}}$ = 950 $s^{-1}$.

Fig. 8 shows the comparison of the distributions of temperatures and volumetric HRRs of $CH_4$ and CR16 counterflow premixed flames at $\Phi = 0.5$ and 0.66. The black and red lines represent the temperature and volumetric HRR, respectively, whereas the solid and dashed lines denote the $CH_4$ and CR16 flames, respectively. The blue dotted line indicates the location of the stagnation plane. Fig. 8(a) presents the results obtained by applying the global strain rate immediately prior to the extinction of the $CH_4$ flame at $\Phi = 0.5$, $a_{\mathrm{g}} = 133\ \mathrm{s}^{-1}$ to both flames. As shown in Fig. 5, under this condition, the $CH_4$ flame is close to its extinction limit, whereas the global extinction strain rate of the CR16 flame is $400\ \mathrm{s}^{-1}$, indicating that it has not yet reached its extinction limit. Therefore, this comparison was conducted to examine how the CR16 flame maintains the temperature and heat release within the reaction zone, whereas the $CH_4$ flame approaches extinction under the same global strain-rate condition. As shown in Fig. 8(a), the CR16 flame exhibits a broader temperature distribution and larger distance between the reaction zones than those of the $CH_4$ flame, and the heat-release reaction zone is sustained farther from the stagnation plane. This indicates that the CR16 flame can maintain chemical reactions over a wider spatial region under the same global strain-rate condition. Fig. 8(b) shows the results at $\Phi = 0.66$, where the $CH_4$ and CR16 flames have identical global extinction strain rates of $a_g = 950\ s^{-1}$, as shown in Fig. 5. In comparison with the $\Phi = 0.5$ condition in Fig. 8(a), both flames experience a higher global strain rate, causing the temperature and heat-release reaction zones to become compressed toward the stagnation plane and the distance between the reaction zones to decrease.

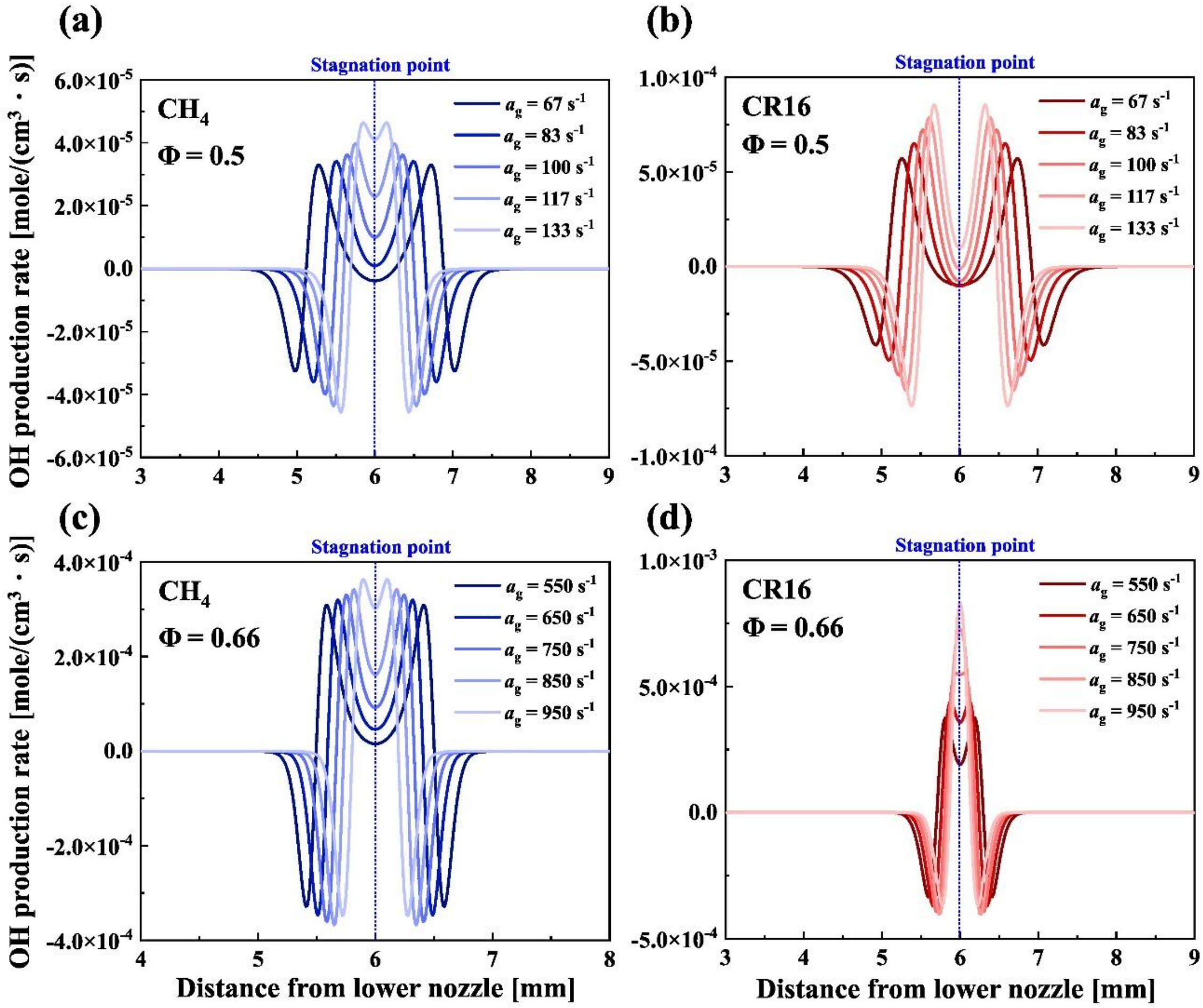

**Fig. 9.** Axial profiles of net OH molar production rates of $CH_4$ and CR16 counterflow premixed flames at varying global strain rates: (a) $CH_4$ and (b) CR16 at Φ = 0.5, and (c) $CH_4$ and (d) CR16 at Φ = 0.66.

Fig. 9 presents the distributions of the net OH molar production rates of $CH_4$ and CR16 flames as functions of the global strain rate at the same equivalence ratios of $\Phi = 0.5$ and 0.66 considered in Fig. 8. OH is a key active radical that undergoes repeated production and consumption during combustion. Positive and negative values of the net OH production rate represent the net production and net consumption of OH, respectively. Accordingly, these results were used to identify the active-radical production–consumption structure within the reaction zone and changes in the location of the chemical-reaction region. For both fuels, increasing the global strain rate shifted the OH production and consumption zones toward the stagnation plane and concentrated them within a narrower spatial region. This behavior results from the compression of the reaction zones toward the stagnation plane under an increased strain. At $\Phi = 0.5$ and $a_\mathrm{g} = 133\ \mathrm{s}^{-1}$, the CR16 flame exhibited a broader OH production–consumption region and higher OH production rate than those of the $CH_4$ flame. This indicates that active-radical production reactions in the CR16 flame were sustained over a wider spatial region under the same strain-rate condition.

At $\Phi = 0.66$, the $CH_4$ flame maintained two distinct OH production zones on either side of the stagnation plane until immediately prior to extinction. In contrast, as the strain rate increased, the two OH production zones in the CR16 flame approached the stagnation plane and ultimately formed a single OH production zone concentrated around the stagnation plane, with a higher OH production rate immediately prior to extinction. These results indicate that the active-radical chemistry within the interflame region of the CR16 flame is sustained up to the vicinity of the stagnation plane. The implications of this behavior were further examined by analyzing the major chemical species, radical production and consumption reactions, and dominant reaction pathways to clarify the relationship between this OH reaction structure and extinction resistance of CR16.

### *3.4. Flame structures*

As discussed above, the temperature, HRR, and net OH production-rate distributions of $CH_4$ and CR16 counterflow premixed flames were compared at $\Phi = 0.5$ and 0.66 to examine the locations of the reaction zones and active-radical reaction structures under variation in the global strain rate. However, to more clearly interpret the differences in the heat-release and temperature distributions observed in flames immediately prior to extinction, it is necessary to analyze the formation and consumption characteristics of the major reactants, products, and intermediate species together with the heat-release contributions of individual reaction pathways. Accordingly, in this section, the reaction rates of major chemical species and reaction-specific heat-release contributions of $CH_4$ and CR16 flames under near-extinction conditions were compared to examine in detail the chemical-reaction structure sustained in the CR16 flame within the interflame region and near the stagnation plane.

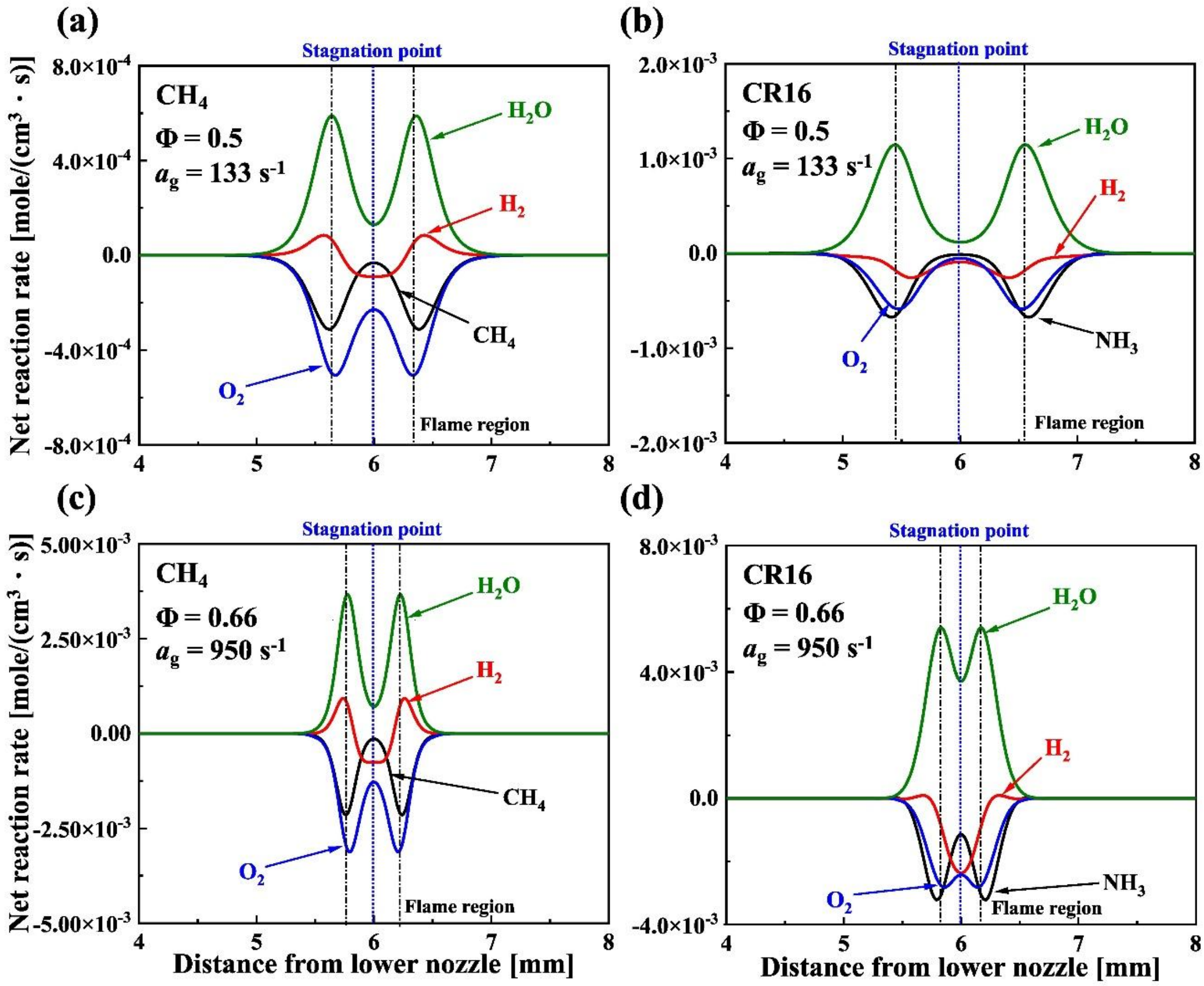

**Fig. 10.** Axial profiles of net molar production rates of major species in $CH_4$ and CR16 counterflow premixed flames at (a, b) $\mathbf{\Phi}$ = 0.5, $\boldsymbol{a}_{\mathbf{g}}$ = 133 s$^{-1}$ and (c, d) $\mathbf{\Phi}$ = 0.66, $\boldsymbol{a}_{\mathbf{g}}$ = 950 s$^{-1}$.

Fig. 10 presents the net molar production-rate distributions of the major chemical species in $CH_4$ and CR16 counterflow premixed flames at $\Phi = 0.5$ and 0.66. The black dash-dotted lines indicate the locations of the maximum HRRs, i.e., the positions of the individual flame zones.

For both fuels, the locations of the maximum $H_2O$ production rates generally coincide with those of the maximum HRRs shown in Fig. 8. This indicates that fuel and oxidizer consumption, together with product formation, occurs actively within the main flame-zone reaction regions. Under the conditions of $\Phi = 0.5$ and $a_{\mathrm{g}} = 133\ \mathrm{s}^{-1}$, shown in Figs. 10(a) and (b), $CH_4$ and $O_2$ were mainly consumed near the locations of the maximum heat release in the $CH_4$ flame, accompanied by $H_2O$ formation. In addition, $H_2$ produced during the $CH_4$ oxidation process was consumed in the region between the flame reaction zone and stagnation plane. In contrast, in the CR16 flame, the consumption of $NH_3$ and $O_2$ mainly occurred near the respective flame-zone locations, whereas $H_2$ exhibited active consumption at locations distinct from those of $NH_3$, with its consumption reactions remaining active up to the vicinity of the stagnation plane.

Similarly, under the conditions of $\Phi = 0.66$ and $a_{\mathrm{g}} = 950\ \mathrm{s}^{-1}$, shown in Figs. 10(c) and (d), the $CH_4$ flame exhibited fuel and $O_2$ consumption and $H_2O$ formation concentrated within the separated reaction zones on either side of the stagnation plane. In contrast, for the CR16 flame, the region of active $NH_3$ consumption was spatially distinct from that of active $H_2$ consumption, and the $H_2$ consumption reactions persisted throughout the interflame region and up to the vicinity of the stagnation plane. This distinct spatial distribution of the major species in the CR16 flame, when compared with that of the $CH_4$ flame, is likely associated with the high diffusivity and preferential-diffusion behavior of $H_2$. It is considered to facilitate the maintenance of reactivity and heat release near the stagnation plane in the CR16 flame.

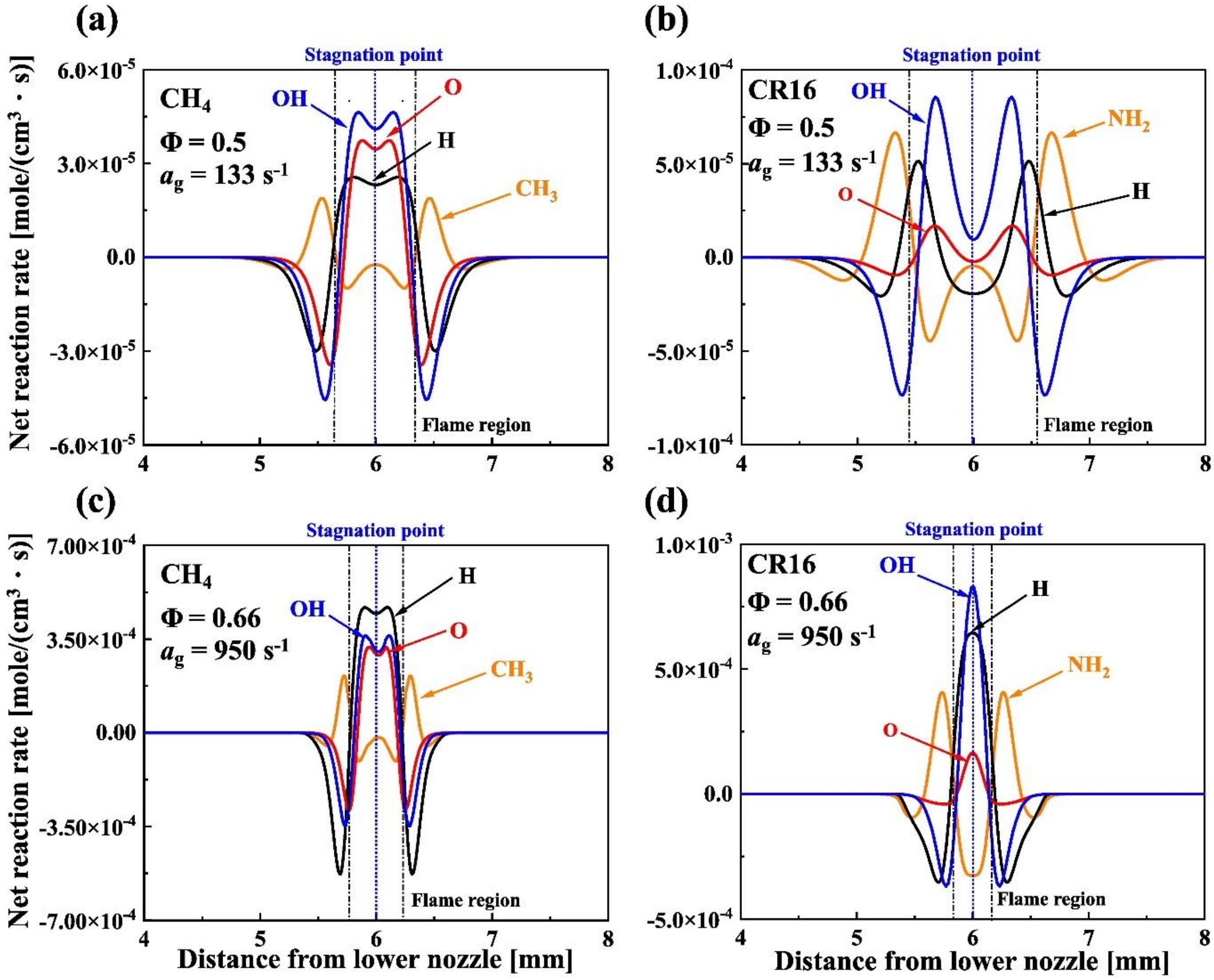

**Fig. 11.** Axial profiles of the net molar production rates of key radicals in $CH_4$ and CR16 counterflow premixed flames at (a, b) $\boldsymbol{\Phi}$ = 0.5, $\boldsymbol{a}_{\mathbf{g}}$ = 133 s$^{-1}$ and (c, d) $\boldsymbol{\Phi}$ = 0.66, $\boldsymbol{a}_{\mathbf{g}}$ = 950 s$^{-1}$.

Fig. 11 shows the distributions of the net molar production rates of major radicals in the $CH_4$ and CR16 counterflow premixed flames at $\Phi = 0.5$ and 0.66. In these distributions, $CH_3$ and $NH_2$ are the major radicals formed during the initial oxidation processes of $CH_4$ and $NH_3$, respectively. These radicals subsequently affect the fuel oxidation and heat-release processes through chain reactions involving active radicals such as H, O, and OH. At $\Phi = 0.5$ and $a_{\mathrm{g}} = 133\ \mathrm{s}^{-1}$, the $CH_4$ flame exhibited $CH_3$ production and consumption reactions, together with H, O, and OH radical reactions, distributed within the flame-zone reaction regions on either side of the stagnation plane. In the CR16 flame, $NH_2$ was the major radical formed during $NH_3$ oxidation, whereas the production and consumption reactions of H, O, and OH were sustained across both the flame reaction zones and the interflame region.

At $\Phi = 0.66$ and $a_g = 950\ s^{-1}$, the H-, O-, and OH-related reactions in both the $CH_4$ and CR16 flames were concentrated within a narrow region near the stagnation plane because of the high global strain rate. In the $CH_4$ flame, the $CH_3$-related reactions and H-, O-, and OH-related reactions were jointly distributed within narrow reaction zones on either side of the stagnation plane. In contrast, in the CR16 flame, the $NH_2$ production and consumption reactions were located on either side of the stagnation plane, whereas H- and OH-related reactions exhibited a more concentrated distribution in the vicinity of the stagnation plane. These results demonstrate that, even at the same global extinction strain rate, the two fuels sustain radical reactions near the stagnation plane through different radical reaction pathways.

Next, the contributions of the major reaction pathways to the heat release were analyzed for the $CH_4$ and CR16 flames. The HRR of each reaction pathway was integrated along the axial direction to quantify the heat-release contribution of each reaction. The major reaction pathways with substantial heat-release contributions under each condition were selected, and the results are summarized in Tables 2–5. The five dominant elementary reactions contributing to heat release were selected based on the absolute magnitude of the integrated reaction-specific heat-release rate (HRR) over the entire axial computational domain. The integrated HRR for each elementary reaction was calculated using Eq. (8):

$$\text{Integrated HRR} = \int_{x_0}^{x_1} \text{HRR}(x)\, dx \qquad (8)$$

where $x_0$ and $x_1$ denote the axial locations of the exits of nozzle 1 and nozzle 2, respectively. Thus, the integration was performed over the entire axial domain between the two opposed nozzle exits. Positive and negative integrated HRR values indicate net exothermic and endothermic contributions, respectively. The five dominant elementary reactions were then identified based on the absolute magnitude of the signed integrated HRR values. These results were used to identify the reaction pathways governing heat release under each condition for the $CH_4$ and CR16 flames and to compare the differences between the two flames from the perspective of extinction resistance.

**Table 2.** Integrated HRR of reaction paths with a large contribution to the HRR for $CH_4$, Φ = 0.5, and $a_g$ = 133 $s^{-1}$.

| Reaction path | Integrated HRR (×$10^4$ J/$m^2$·s) |
|---|---|
| $O + CH_3 \rightleftharpoons H + CH_2O$ (R10) | 1.98 |
| $H + O_2 \rightleftharpoons O + OH$ (R38) | −1.97 |
| $OH + HO_2 \rightleftharpoons O_2 + H_2O$ (R87) | 1.39 |
| $OH + CO \rightleftharpoons H + CO_2$ (R99) | 1.59 |
| $HCO + O_2 \rightleftharpoons HO_2 + CO$ (R168) | 1.56 |

**Table 3.** Integrated HRR of reaction paths with a large contribution to the HRR for $CH_4$, Φ = 0.66, and $a_g$ = 950 $s^{-1}$.

| Reaction path | Integrated HRR (×$10^4$ J/$m^2$·s) |
|---|---|
| $O + CH_3 \rightleftharpoons H + CH_2O$ (R10) | 8.58 |
| $H + O_2 \rightleftharpoons O + OH$ (R38) | −7.37 |
| $OH + CO \rightleftharpoons H + CO_2$ (R99) | 4.84 |
| $HCO + O_2 \rightleftharpoons HO_2 + CO$ (R168) | 4.33 |
| $O + CH_3 \rightleftharpoons H + H_2 + CO$ (R284) | 5.62 |

**Table 4.** Integrated HRR of reaction paths with a large contribution to the HRR for CR16, Φ = 0.5, and $a_g$ = 133 $s^{-1}$.

| Reaction path | Integrated HRR (×$10^4$ J/$m^2$·s) |
|---|---|
| $H_2 + OH \rightleftharpoons H_2O + H$ (R3) | 2.12 |
| $H + O_2 \rightleftharpoons OH + O$ (R38) | −3.47 |
| $NH_2 + NO \rightleftharpoons N_2 + H_2O$ (R277) | 5.02 |
| $NH_3 + OH \rightleftharpoons NH_2 + H_2O$ (R282) | 2.44 |
| $HNO + OH \rightleftharpoons NO + H_2O$ (R289) | 4.45 |

**Table 5.** Integrated HRR of reaction paths with a large contribution to the HRR for CR16, Φ = 0.66, and $a_g$ = 950 $s^{-1}$.

| Reaction path | Integrated HRR (×$10^4$ J/$m^2$·s) |
|---|---|
| $H_2 + OH \rightleftharpoons H_2O + H$ (R3) | 8.13 |
| $H + O_2 \rightleftharpoons OH + O$ (R38) | −10.2 |
| $NH_3 + OH \rightleftharpoons NH_2 + H_2O$ (R282) | 6.86 |
| $HNO + H \rightleftharpoons NO + H_2$ (R288) | 8.69 |
| $HNO + OH \rightleftharpoons NO + H_2O$ (R289) | 7.05 |

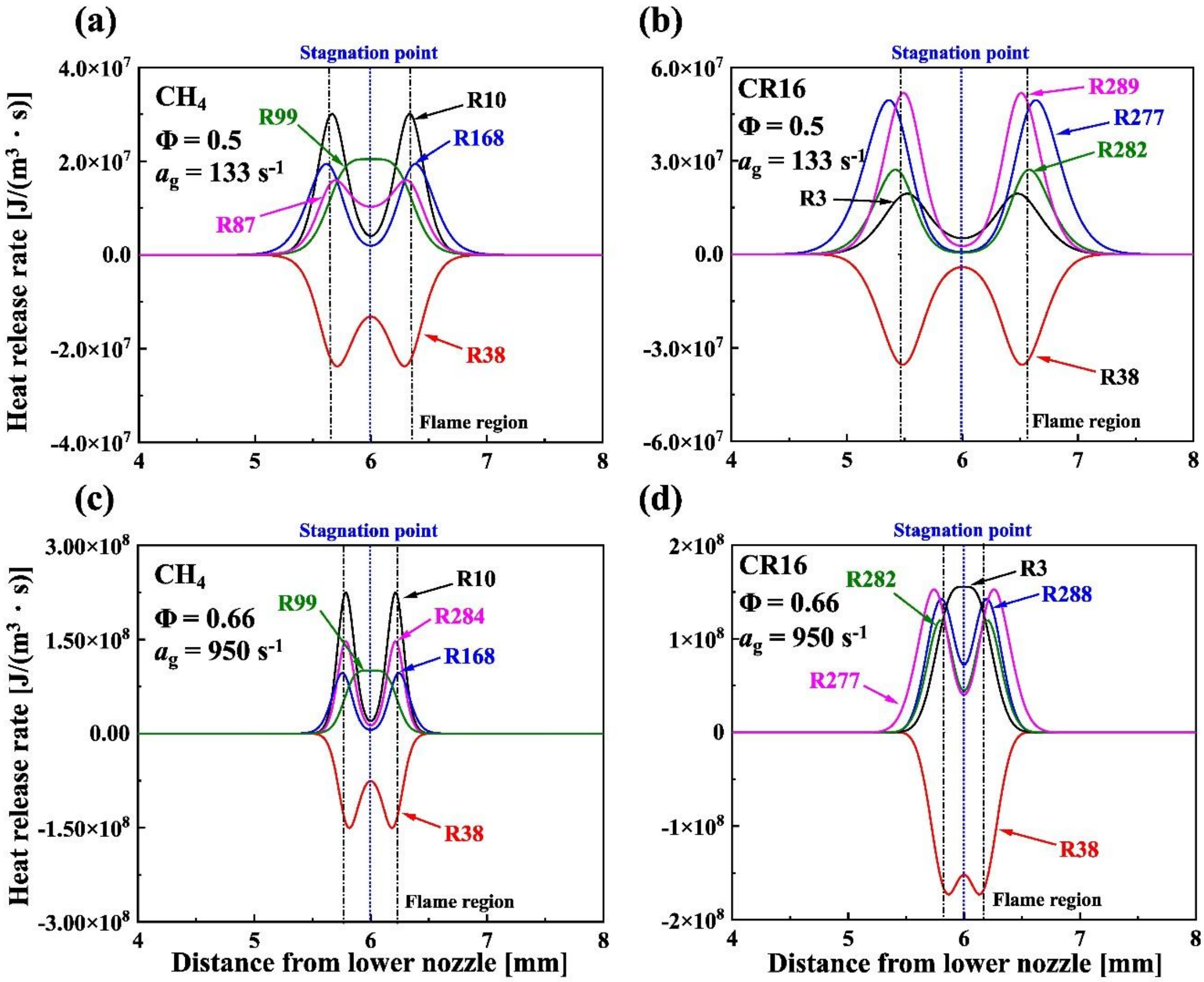

**Fig. 12.** Axial profiles of heat-release rate contributions of the five dominant elementary reactions in $CH_4$ and CR16 counterflow premixed flames at (a, b) $\boldsymbol{\Phi}$ = 0.5, $\boldsymbol{a_g}$ = 133 $s^{-1}$ and (c, d) $\boldsymbol{\Phi}$ = 0.66, $\boldsymbol{a_g}$ = 950 $s^{-1}$.

Fig. 12 presents the distributions of the HRR of the reactions summarized in Tables 2–5. Positive and negative values represent exothermic and endothermic reactions, respectively. Therefore, Fig. 12 shows not only the overall heat-release contributions of the major reactions but also the distributions of the reactions contributing to heat absorption. As indicated by the values summarized in Tables 2–5, a substantial endothermic reaction distribution associated with $H + O_2 \rightleftharpoons O + OH$(R38) was observed under all conditions. R38 is a representative chain-branching reaction that produces H, O, and OH radicals. In addition, although it is not directly exothermic, it plays an important role in the formation and maintenance of radicals within the flame [36]. First, at $\Phi = 0.5$ and $a_g = 133\ s^{-1}$, the major exothermic reactions in the $CH_4$ flame were $O + CH_3 \rightleftharpoons H + CH_2O$ (R10), $OH + CO \rightleftharpoons H + CO_2$ (R99), and HCO oxidation reactions. Their heat-release contributions were mainly concentrated within the flame-zone reaction regions on either side of the stagnation plane. In contrast, the major exothermic pathways in the CR16 flame included $H_2 + OH \rightleftharpoons H + H_2O$ (R3), $NH_3 + OH \rightleftharpoons NH_2 + H_2O$ (R282), $NH_2 + NO \rightleftharpoons N_2 + H_2O$ (R277), and HNO-related reactions. In particular, hydrogen oxidation and nitrogen-related reactions in the CR16 flame contributed to heat release not only within the flame reaction zones but also throughout the interflame region adjacent to the stagnation plane.

At $\Phi = 0.66$ and $a_g = 950\ s^{-1}$, the major reactions in both flames were compressed into a narrow interflame region near the stagnation plane because of the high global strain rate. In the $CH_4$ flame, $CH_3$ oxidation and CO/HCO oxidation pathways provided the major heat release. In contrast, in the CR16 flame, hydrogen oxidation (R3), $NH_3$ activation (R282), and HNO-involving reactions jointly contributed to the major exothermic pathways. In particular, as shown in Table 5, the largest heat-release contribution in the CR16 flame was from $HNO + H \rightleftharpoons NO + H_2$ (R288), whereas R3 and R282 also exhibited substantial exothermic contributions. These results show that, unlike the $CH_4$ flame, which is dominated by a single hydrocarbon-oxidation pathway, the CR16 flame sustains chemical reactions under high-strain conditions through multiple coupled pathways involving hydrogen oxidation and nitrogen-related chemistry.

This reaction structure can be interpreted as the chemical basis enabling the CR16 flame to maintain reactivity and heat release within the interflame region, even under high-strain conditions. Owing to its greater flame thickness and lower laminar-flame speed, CR16 has a longer flame time than that of $CH_4$. In addition, the coupled hydrogen-oxidation and nitrogen-related reaction pathways enable chemical reactions to persist under high-strain conditions. Therefore, the combined effects of the longer flame time and sustained chemical reactivity are considered to result in the higher RSIB of the CR16 flame relative to that of the $CH_4$ flame.

# 4. Conclusion

In this study, the extinction behaviors of lean counterflow premixed flames fueled by $CH_4$ and CR16, a 16% cracked ammonia surrogate fuel with a high RSIB value and favorable cracking economics (72% $NH_3$/21% $H_2$/7% $N_2$), were compared experimentally and numerically. The flame structure and major reaction characteristics were also analyzed. The main conclusions are as follows.

The CR16 flame exhibited a higher extinction strain rate than that of the $CH_4$ flame under the lean conditions of $\Phi = 0.5$–$0.65$, confirming its greater resistance to stretch-induced extinction. In particular, at $\Phi = 0.66$, the two flames exhibited the same extinction strain rate of $950\ \mathrm{s}^{-1}$; however, the CR16 flame showed a higher RSIB value, indicating superior extinction-resistance characteristics. Freely propagating flame simulations showed that the CR16 flame had a lower laminar-flame speed and greater flame thickness than those of the $CH_4$ flame, resulting in a longer flame time. This extended flame time is considered a major factor associated with the higher RSIB and enhanced extinction resistance of the CR16 flame under lean conditions.

A comparison of the near-extinction flame structures showed that the CR16 flame maintained higher temperatures and HRRs than those of the $CH_4$ flame, suggesting an enhanced resistance to extinction. In particular, owing to the preferential diffusion of $H_2$, $H_2$-consumption reactions remained active even near the stagnation plane.

A comparison of the major radical reactions showed that the production rates of the H, O, and OH radicals were higher in the CR16 flame than in the $CH_4$ flame, indicating their contribution to radical maintenance and activation of chain reactions within the flame. In addition, the analysis of the distributions of the major reaction pathways confirmed that $NH_2 + NO \rightleftharpoons N_2 + H_2O$, $HNO + OH \rightleftharpoons NO + H_2O$, and $H_2 + OH \rightleftharpoons H_2O + H$ acted as the major exothermic pathways in the CR16 flame. In contrast, although $H + O_2 \rightleftharpoons O + OH$ is an endothermic reaction, it was shown to play an important role in maintaining the flame reactivity as a chain-branching reaction that generates H, O, and OH radicals.

## Acknowledgment

This study was supported by the National Research Foundation of Korea (NRF) grant funded by the Korean Government (Ministry of Science and ICT) (No. RS-2025-16069888), Korea Evaluation Institute of Industrial Technology (KEIT), and Ministry of Trade, Industry & Energy (MOTIE) of the Republic of Korea (RS-2025-07982969, Development of real-time analysis of hydrogen mixed gas and flame diagnostic combustion control technology). Professor Masri is supported by the Australian Research Council.